\documentclass{article}

\usepackage[preprint,nonatbib]{neurips_2024}

\usepackage[utf8]{inputenc} 
\usepackage[T1]{fontenc}    
\usepackage[hidelinks,colorlinks=true,linkcolor=blue,citecolor=blue]{hyperref}
\usepackage{url}            
\usepackage{booktabs}       
\usepackage{amsfonts}       
\usepackage{nicefrac}       
\usepackage{microtype}      
\usepackage{xcolor}         
\usepackage{titlesec}
\usepackage{graphicx}
\usepackage{wrapfig}
\usepackage{bm}
\usepackage{amsmath}
\usepackage{setspace}
\usepackage{amsfonts}
\usepackage{amssymb}
\usepackage{amsthm} 
\usepackage{multicol}
\usepackage{graphicx}
\usepackage{enumerate}

\usepackage{xspace}
\usepackage{xspace}
\usepackage{multirow}
\usepackage{pythonhighlight}
\usepackage{algorithm}
\usepackage{algorithm,algpseudocode}
\usepackage{longtable}
\usepackage{pdfpages}
\usepackage{xr}
\usepackage{cite}

\def\given{\mid}

\def\set#1{\mathcal{#1}}

\def\numMethods{26\xspace}
\def\numNetworks{90\xspace}

\def\method#1{\texttt{#1}\xspace}
\makeatletter

\makeatletter
\renewcommand\subsubsection{\@startsection{subsubsection}{3}{\z@}%
                                     {-0.5ex\@plus -1ex \@minus -.2ex}%
                                     {-1.5ex \@plus -.2ex}
                                     {\normalfont\normalsize\bfseries}}
\makeatother

\title{Implicit degree bias in the link prediction task}

\author{%
  Rachith Aiyappa$ ^1$, Xin Wang$ ^3$, Munjung Kim$ ^1$, Ozgur Can Seckin$ ^1$, \AND Jisung Yoon$ ^2$,  Yong-Yeol Ahn$ ^1$, Sadamori Kojaku$ ^3$ \AND
  \\ \vspace{-1em}
  \begin{minipage}{15em}
    \centering
     $ ^1$Center for Complex Networks and Systems Research, Luddy School of Informatics, Computing, and Engineering, Indiana University, Bloomington, IN, USA
  \end{minipage}
  \And
  \\\vspace{-1em}
  \begin{minipage}{10em}
    \centering
    $ ^2$KDI School of Public Policy and Management, Sejong City, South Korea
  \end{minipage}
  \And
  \\ \vspace{-1em}
  \begin{minipage}{11em}
    \centering
    $ ^3$Department of Systems Science and Industrial Engineering, Binghamton University, Binghamton, NY, USA
  \end{minipage}
  \And
  \\ \vspace{-1em}
  \begin{minipage}{\hsize}
    \centering
    \texttt{\{racball,munjkim,oseckin,yyahn\}@iu.edu, jisung.yoon92@gmail.com,\{xwang314,skojaku\}@binghamton.edu}
  \end{minipage}
}

\begin{document}

\maketitle

\begin{abstract}
Link prediction---a task of distinguishing actual hidden edges from random unconnected node pairs---is one of the quintessential tasks in graph machine learning.
Despite being widely accepted as a universal benchmark and a downstream task for representation learning, the validity of the link prediction benchmark itself has been rarely questioned.
Here, we show that the common edge sampling procedure in the link prediction task has an implicit bias toward high-degree nodes and produces a highly skewed evaluation that favors methods overly dependent on node degree, to the extent that a ``null'' link prediction method based solely on node degree can yield nearly optimal performance.
We propose a degree-corrected link prediction task that offers a more reasonable assessment that aligns better with the performance in the recommendation task.
Finally, we demonstrate that the degree-corrected benchmark can more effectively train graph machine-learning models by reducing overfitting to node degrees and facilitating the learning of relevant structures in graphs.
\end{abstract}

\section{Introduction}
\label{sec:introduction}
%
Standardized benchmarks such as ImageNet~\cite{deng2009imagenet,krizhevsky2012imagenet} and SQuAD~\cite{rajpurkar-etal-2016-squad,rajpurkar-etal-2018-know} play a pivotal role in guiding the evolution of machine learning, creating a competitive environment that propels innovation by setting clear, measurable goals.
A core benchmark for graph machine learning is the link prediction task, the task of identifying missing edges in a graph, with diverse applications including the recommendations of friends and contents~\cite{kunegis2009learning,wang2014link,huang2005link,menon2011link}, knowledge discoveries~\cite{sun2019rotate,bordes2013translating}, and drug development~\cite{abbas2021application,breit2020openbiolink,you2019predicting,wang2015overactive,crichton2018neural,yue2020graph,ali2019biokeen}.
The link prediction benchmarks have been serving as a crucial benchmark that facilitates quantitative evaluations and drives the advancement of graph machine learning techniques~\cite{ghasemian2020stacking,libenSocialNetworks2003,mara2020benchmarking,breit2020openbiolink,yue2020graph,ali2019biokeen,narayanan2011link}.

Despite its significant role in graph machine learning, the link prediction benchmark itself has rarely been examined critically for its effectiveness, reliability, and potential biases.
A widely-used link prediction benchmark evaluates methods by their ability to classify pairs of nodes as either connected or unconnected~\cite{kunegis2009learning,ghasemian2020stacking,mara2020benchmarking}.
The connected node pairs (i.e., edges) are randomly sampled from existing edges as the hidden positive set, and an equal number of node pairs are randomly chosen from unconnected node pairs.
Existing criticisms mainly consider the disconnect from real-world scenarios.
For instance, unconnected node pairs are far more common than connected node pairs because of the sparsity of edges in graphs~\cite{newman2018network,barabasiNetworkScience2016}, which can lead to biased performance evaluations~\cite{li2024evaluating,menand2024link,yang2015evaluating,huang2023link,wang2021pairwise}.
Additionally, while the link prediction benchmark tests methods by classifying a predefined set of potential edges, the practical link prediction task involves identifying the potential edges from the entire graph.
Despite the misalignment with real-world scenarios, high benchmark performance is often regarded as a testament to successful learning in graph machine learning~\cite{ghasemian2020stacking,zhang2018link,breit2020openbiolink,crichton2018neural,yue2020graph,ali2019biokeen,groverNode2vecScalableFeature2016,ou2016asymmetric,goyal2018graph,cai2021line}.

Here, we argue that the standard link prediction benchmark has a fundamental and severe bias that favors methods that exploit node degree information (i.e., the number of edges a node has).
The degree bias arises from the sampling of edges used for performance evaluations, i.e.,
when sampling an edge from a graph uniformly at random, a node with $k$ edges is $k$ times more likely to be selected than a node with a single edge ($k=1$).
Meanwhile, the negative set of edges is randomly sampled from unconnected node pairs, which do not have this degree bias.
As a result, we have a distinct and useful feature (degree) for the methods that can be leveraged without understanding any non-trivial structural features of the graph.
We show that this degree bias is so profound that a ``null'' link prediction method based solely on node degree can yield nearly optimal performance, questioning the usefulness of the link prediction benchmark even as a general objective for graph machine learning and raising awareness on the need for being more intentional and careful about what the evaluation tasks themselves actually evaluate.

To address this bias, we propose a degree-corrected link prediction benchmark that samples the unconnected node pairs with the same degree bias.
We demonstrate that the degree-corrected link prediction benchmark can better capture the performance of algorithms in the recommendation task.
Moreover, we show that the degree-corrected benchmark trains graph neural networks more effectively by reducing overfitting to node degrees, thereby improving the learning of community structure in graphs.

%
%
\section{Design flaw of the link prediction benchmark}
\label{sec:link_prediction}

\subsubsection*{Preliminary}\label{sec:preliminary}

We focus on the unweighted, undirected graph $G = (\set{V}, \set{E})$, where $\set{V}$ is the set of nodes and $\set{E}$ is the set of edges.
We assume that $G$ has no self-loops, no multiple edges, and is highly sparse ($|\set{E}| \ll |\set{V}|^2$), which is a common characteristic of real-world graphs~\cite{newman2018network,barabasiNetworkScience2016}.
Degree $k_i$ of a node $i \in \mathcal{V}$ is the number of edges emanating from node $i$.
We use $\sim$ to denote proportional relationships.
We exclude node attributes from data, if present, to ensure consistency across all graph-based link prediction methods.


\begin{figure}
    \centering
    \includegraphics[width=\hsize]{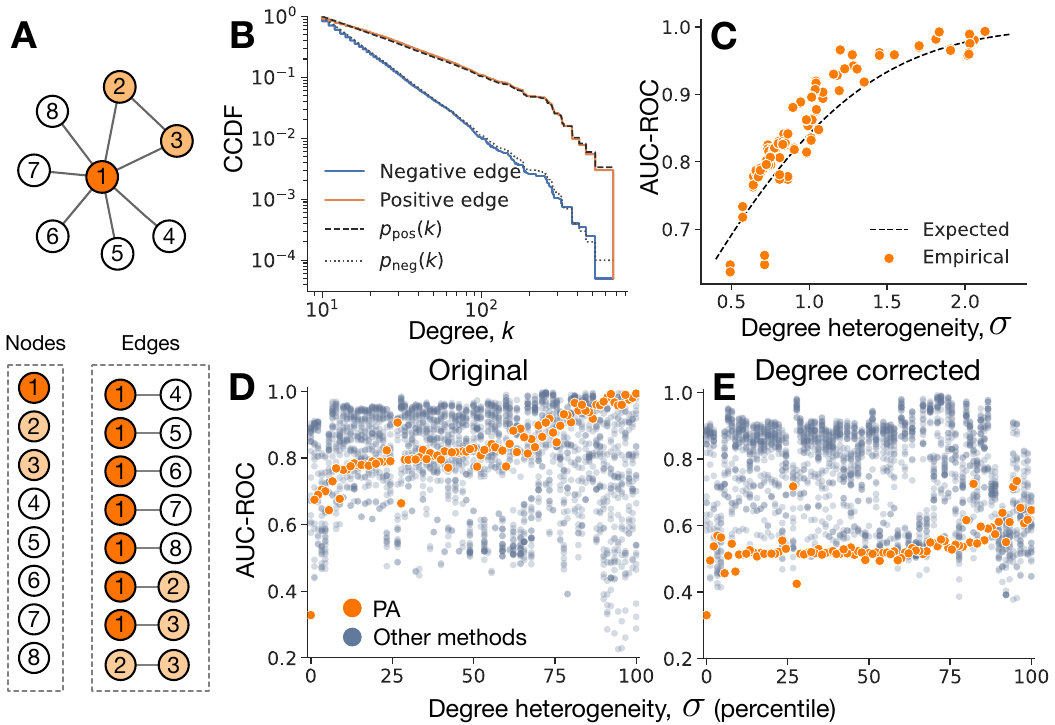}
    \caption{
        Illustration of the degree bias in the link prediction benchmark.
        {\bf A}: A node with degree $k$ appears $k$ times in the edge list, making it $k$ times more likely to be sampled as a positive edge than a node with degree 1.
        {\bf B}: The degree distribution of the nodes in the positive and negative edges sampled from a Price graph of $N=10^5$ nodes and $M=10^6$ edges.
        The y-axis, ``CCDF'', denotes the complementary cumulative distribution function, representing the probability that a node's degree is at least $k$. Dashed lines illustrate the relationship described by Eq.~\eqref{eq:pos-neg-degree-dist}.
        {\bf C}: The AUC-ROC score for the Preferential Attachment (PA) method on empirical graphs, with the dashed line indicating the expected AUC-ROC based on the ``null'' link prediction method that uses only the node degree information (Eq.~\eqref{eq:auc-roc-log-normal}).
        {\bf D}: AUC-ROC of \numMethods methods across \numNetworks graphs.
        PA outperforms 54\% of the method on average across \numNetworks graphs and outperforms most other methods in the most heterogeneous graphs.
        {\bf E}: AUC-ROC of the same methods for the degree-corrected benchmark, showing that nearly 86\% of the methods perform better than PA on average.
    }
    \label{fig:bias-illustration}
\end{figure}

\subsubsection*{Link prediction benchmark}
\label{sec:standard_benchmark}

The standard procedure for link prediction benchmarks is outlined as follows~\cite{kunegis2009learning,zhang2018link,breit2020openbiolink,crichton2018neural,yue2020graph,ali2019biokeen,groverNode2vecScalableFeature2016,ou2016asymmetric,goyal2018graph,cai2021line}.
First, we randomly sample a fraction $\beta$ of edges from the edge set $\set{E}$ as \emph{positive edges}.
Second, we randomly sample an equal number of unconnected node pairs with replacement from the node set $\set{V}$ as \emph{negative edges}.
We resample the negative edges if they form a loop or are already in the positive or test edges.
Third, each node pair $(i,j)$ is scored by a link prediction method, where a higher score $s_{ij}$ indicates a higher likelihood of an edge existing between the nodes.
Fourth, the effectiveness of a method is evaluated using the Area Under the Receiver Operating Characteristic Curve (AUC-ROC), which represents the probability that the method gives a higher score to a positive edge than a negative edge.
While alternative benchmark designs use different evaluation metrics or sampling strategies for negative edges~\cite{li2024evaluating,menand2024link,yang2015evaluating,huang2023link,wang2021pairwise} (Section~\ref{sec:discussion}), this outlined procedure has been widely used~\cite{kunegis2009learning,zhang2018link,breit2020openbiolink,crichton2018neural,yue2020graph,ali2019biokeen,groverNode2vecScalableFeature2016,ou2016asymmetric,goyal2018graph,ghasemian2020stacking}.

\subsubsection*{Sampling bias due to node degree}
\label{sec:design_flaw}

A well-known, counterintuitive fact about graphs is that a uniform random sampling of edges results in a \emph{biased} selection of nodes in terms of degree~\cite{feld1991your,barthelemy2004velocity,christakis2010social,kojakuEffectivenessBackwardContact2021,kojaku2021residual2vec}.
The bias arises because a node with $k$ edges appears $k$ times in the edge list and thus $k$ times more likely to be chosen than a node with $k'=1$ edge (e.g., node 1 and 4 in Fig.~\ref{fig:bias-illustration}A).
Consequently, for a graph with degree distribution $p(k)$, the nodes in the positive edges have degree distribution $p_{\text{pos}}(k)$ proportional to $\sim k\cdot p(k)$.
By normalizing $k\cdot p(k)$, the degree distribution of the positive edges is given by
\begin{align}
    \label{eq:pos-neg-degree-dist}
    p_{\text{pos}}(k) = \frac{1}{\sum_{\ell} \ell p(\ell)} k \cdot p(k) = \frac{1}{\langle k \rangle} k\cdot p(k),
\end{align}
where $\langle k \rangle$ is the average degree of the graph.
On the other hand, the nodes in the negative edges are uniformly sampled from the node list (i.e., $\set{V}$) and thereby have degree distribution $p_{\text{neg}}(k)$ identical to $p(k)$ (i.e., $p_{\text{neg}}(k) = p(k)$).

We demonstrate the degree bias using the Price graph of $N=10^5$ nodes and $M=10^6$ edges, with the power-law degree distribution approximating $p(k)\propto k^{-3}$ (Fig.~\ref{fig:bias-illustration}B).
We sample $\beta = 0.25$ of the edges uniformly at random from $\set{E}$ as positive edges, together with the same number of unconnected node pairs by uniformly sampling nodes from $\set{V}$.
The degree distributions for nodes in the positive and negative edges closely follow $p_{\text{pos}}(k)$ and $p_{\text{neg}}(k)$, respectively, demonstrating the sampling bias due to node degree.
We note that the degree bias is not specific to the Price graph but appears for any graph with non-uniform degree distribution.

\subsubsection*{Impact of degree bias on the link prediction benchmark}

We demonstrate the impact of degree bias on link prediction benchmark (Fig.\ref{fig:bias-illustration}D).
We evaluated \numMethods link prediction methods across \numNetworks graphs. These methods include 7 network topology-based methods (e.g., Common Neighbors (\method{CN})~\cite{libenSocialNetworks2003}), 13 graph embedding methods (e.g., Laplacian EigenMap (\method{EigenMap})~\cite{belkinLaplacianEigenmapsDimensionality2003}), 2 network models (e.g., Stochastic Block Model (e.g., \method{SBM})~\cite{fortunatoCommunityDetectionGraphs2010}), and 4 graph neural networks (GNNs)  (e.g., Graph Convolutional Network (\method{GCN})~\cite{kipf2017semi}). Detailed descriptions of the methods and graphs are available in SI Section~1.
We set the fraction of test edges to $\beta = 0.25$ and repeat the experiment 5 times.
We quantify the heterogeneity $\sigma$ of node degree by fitting a log-normal distribution to $p(k)$ and calculating its variance parameter $\sigma$. We will show that $\sigma$ is a good indicator of the impact of degree bias in Section~\ref{sec:pa_is_best}.

We focus on the Preferential Attachment (\method{PA}) link prediction method, which calculates the prediction score $s_{ij} = k_i k_j$ using only node degrees.
Although \method{PA} is a crude method that neglects key predictive features such as the number of common neighbors and shortest distance~\cite{li2024evaluating,menand2024link,lichtnwalter2012link,zhang2018link,mao2023revisiting}, it still outperforms more than half of the advanced methods with an average AUC-ROC of 0.83 (ranked 13th out of \numMethods methods; see Fig.~\ref{fig:uniform-vs-biased}A).
\method{PA} performs better as the heterogeneity of node degrees increases.
The outperformance of \method{PA} is attributed to the degree bias, where the positive edges are more likely to be formed by nodes with high degree and thereby are easily distinguishable from the negative edges.
As a result, the current benchmark design favors methods that make predictions based largely on node degrees.

\subsubsection*{Theoretical analysis}
\label{sec:pa_is_best}

Many empirical graphs exhibit heterogeneous degree distributions, with a few nodes having an exceptionally large degree and most having a small degree.
The heterogeneous degree distributions are often characterized by power-law degree distribution $p(k) \propto k^{-\alpha}$ with $\alpha \in (2,3]$ (i.e., scale-free networks)~\cite{albert2002statistical,barabasi2003scale,holme2019rare,voitalov2019scale} or log-normal distributions~\cite{artico2020rare,broido2019scale}.
While the power-law and log-normal distributions are both continuous, they are often used to characterize the discrete degree distribution and provide approximations~\cite{artico2020rare,broido2019scale,clauset2009power,radicchi2008universality,johnson1995continuous,redner2005citation}.
We show that the AUC-ROC for \method{PA} reaches the nearly maximum under log-normal distributions with heterogeneous node degrees. See SI Section~2.2 for the case of power-law distributions.

Let us consider a general degree distribution $p(k)$ without restricting ourselves to log-normal distributions.
The AUC-ROC has a probablistic interpretation~\cite{hand2009measuring}: it is the probability that the score $s^+$ for the positive edges is larger than the score $s^-$ for the negative edges.
Recalling that \method{PA} computes $s_{ij} = k_i k_j$, the AUC-ROC for \method{PA} is given by
\begin{align}
    \text{AUC-ROC} = P(s_{i^{-},j^{-}} \leq s_{i^{+},j^{+}}) = P(k_{i^{-}} k_{j^{-}} \leq k_{i^{+}} k_{j^{+}}),
    \label{eq:auc-roc}
\end{align}
where $i^{\pm}$ and $j^{\pm}$ represent the nodes of the positive and negative edges, respectively.
Now, let us assume that $p(k)$ follows the log-normal distribution, $\text{LogNorm}(k \given \mu, \sigma^2)$, given by~\cite{hand2009measuring}:
\begin{align}
    p(k) = \text{LogNorm}(k \given \mu, \sigma^2) = \frac{1}{\sqrt{2\pi} \sigma k} \exp\left[ -\frac{(\ln k-\mu)^2}{2\sigma^2} \right],
    \label{eq:log-normal}
\end{align}
where $\mu$ and $\sigma$ are the parameters of the log-normal distribution.
The mean of the log-normal degree distribution is $\langle k \rangle = \exp(\mu + \sigma^2 / 2)$~\cite{hand2009measuring}.
By substitutiong Eq.~\eqref{eq:log-normal} into Eq.~\eqref{eq:pos-neg-degree-dist}, we derive the
degree distribution of nodes in the positive edges as:
\begin{align}
    p_{\text{pos}}(k) &= \frac{k}{\langle k \rangle}  \frac{1}{\sqrt{2\pi} \sigma k} \cdot \exp\left[ -\frac{(\ln k-\mu)^2}{2\sigma^2} \right] \nonumber \\
    &= \frac{1}{\sqrt{2\pi} \sigma k} \exp\left[ -\frac{1}{2\sigma^2}\left( (\ln k)^2 -2\mu \ln k + \mu^2\right) +\ln k - \mu - \sigma^2 / 2\right] \nonumber \\
    &= \frac{1}{\sqrt{2\pi} \sigma k} \exp\left[ -\frac{1}{2\sigma^2}\left( (\ln k)^2 -2(\mu + \sigma^2)\ln k + \mu^2+2\mu \sigma^2 + \sigma^4 \right) \right] \nonumber \\
    &= \frac{1}{\sqrt{2\pi} \sigma k} \exp\left[ -\frac{(\ln k-\mu-\sigma^2)^2}{2\sigma^2} \right]
    = \text{LogNorm}(k \given \mu + \sigma^2, \sigma^2).
    \label{eq:pos-degree-dist-log-normal}
\end{align}
Equation~\eqref{eq:pos-degree-dist-log-normal} indicates that the degree distribution for nodes in the positive edges also follows a log-normal distribution, parameterized by $\mu + \sigma^2$ and $\sigma$.

We derive the AUC-ROC for \method{PA} by leveraging a unique characteristic of log-normal distributions, i.e., the logarithm $\ln k$ of log-normally-distributed degree $k$ follows a normal distribution with mean $\mu$ and variance $\sigma^2$, i.e.,
\begin{align}
    P(\ln k) = \text{Norm}(k \given \mu, \sigma^2),\;\text{where}\;\; \text{Norm}(k \given \mu, \sigma^2) = \frac{1}{\sqrt{2\pi} \sigma} \exp\left[ -\frac{(k-\mu)^2}{2\sigma^2} \right].
\end{align}
We assume no degree assortativity in the graph, where $P(k^+_i,k^+_j) = P(k_i)P(k_j)$. Although empirical graphs often exhibit degree assortativity, our results indicate that it does not significantly impact the AUC-ROC (SI Section~2.1).
For the negative edges, the distribution for $\ln s_{i^-,j^-} = \ln k^- _i + \ln k^- _j$ follows a normal distribution with mean $2\mu$ and variance $2\sigma^2$, as the sum of independent normal variables also forms a normal distribution with additive means and variances~\cite{bishop2006pattern}, i.e.,
\begin{align}
    P(\ln s_{i^-,j^-}) = \text{Norm}\left(\ln s_{i^-,j^-} \given 2\mu, 2\sigma^2\right).
    \label{eq:lns_neg}
\end{align}
For the positive edges, the degree distribution also follows a log-normal distribution (Eq.~\eqref{eq:pos-degree-dist-log-normal}). Thus, the distribution for $\ln s_{i^+,j^+} = \ln k^+ _i + \ln k^+ _j$ is given by
\begin{align}
    P(\ln s_{i^+,j^+}) = \text{Norm}\left(\ln s_{i^+,j^+} \given 2\mu + 2\sigma^2, 2\sigma^2\right).
    \label{eq:lns_pos}
\end{align}
By using Eqs.~\eqref{eq:auc-roc}, \eqref{eq:lns_neg}, and \eqref{eq:lns_pos}, we have
\begin{align}
    \text{AUC-ROC} &= P(\ln s^- < \ln s^+) \nonumber \\
    &= \int_{-\infty} ^{\infty} \text{Norm}(x^- \given 2\mu, 2\sigma^2) \left[1-\int^{x^-} _{-\infty} \text{Norm}(x^+ \given 2\mu + 2\sigma^2, 2\sigma^2) \text{d}x^+ \right] \text{d}x^- \nonumber \\
    &= 1 - \int_{-\infty} ^{\infty} \text{Norm}(x^- \given 2\mu, 2\sigma^2) \int^{x^-} _{-\infty} \text{Norm}(x^+ \given 2\mu + 2\sigma^2, 2\sigma^2) \text{d}x^+ \text{d}x^- \label{eq:auc-roc-log-normal-intermediate}
\end{align}
We reparameterize Eq.~\eqref{eq:auc-roc-log-normal-intermediate} by using $z^\pm = \frac{x^\pm - 2\mu}{\sqrt{2}\sigma}$. Noting that $\text{Norm}(x^- \given 2\mu, 2\sigma^2) \cdot \sqrt{2}\sigma = \text{Norm}(z^- \given 0, 1)$ and ${\rm d}x^\pm = (\sqrt{2}\sigma){\rm d}z^\pm $, we have
\begin{align}
    P(\ln s^- < \ln s^+)
    &= 1 - \int_{-\infty} ^{\infty}(2\sigma^2)  \text{Norm}(x^- \given 2\mu, 2\sigma^2) \int^{x^-} _{-\infty} \text{Norm}(x^+ \given 2\mu + 2\sigma^2, 2\sigma^2) \cdot \text{d}z^+ \text{d}z^- \nonumber \\
    &= 1 - \int_{-\infty} ^{\infty} \text{Norm}\left(z^- \given 0, 1\right) \int^{z^-} _{-\infty} \text{Norm}\left(z^+ -\sqrt{2}\sigma \given 0,1\right) \text{d}z^+ \text{d}z^- \nonumber \\
    &= 1 - \int_{-\infty} ^{\infty} \text{Norm}(z^- \given 0, 1) \Phi\left( z^- -\sqrt{2}\sigma\right) \rm{d}z,
    \label{eq:auc-roc-log-normal}
\end{align}
where $\Phi(z^-)$ is the cumulative distribution function for the standard normal distribution, i.e., $\Phi(z^-) = \int_{\infty} ^{z^-} \text{Norm}(y \given 0, 1){\rm d}y$.
Equation~\eqref{eq:auc-roc-log-normal} suggests the key behavior of AUC-ROC for \method{PA}.
The AUC-ROC for \method{PA} is an increasing function of the parameter $\sigma$ of the log-normal distribution (Fig.~\ref{fig:bias-illustration}C).
The parameter $\sigma$ of the log-normal distribution controls the spread of the distribution, with larger $\sigma$ resulting in a more fat-tailed distribution.
While our assumptions about the log-normal degree distribution and degree assortativity may not always align with real-world data, Eq.~\eqref{eq:auc-roc-log-normal} still effectively captures the AUC-ROC behavior for \method{PA} (Fig.~\ref{fig:bias-illustration}C). Further analysis of power-law distributions is described in SI Section~2.2.

This theoretical result highlights the significant issue with the current link prediction benchmark: a high benchmark performance can be achieved by only learning node degrees, posing the question of whether the link prediction benchmark is an effective objective of graph machine learning.

\section{The degree-corrected link prediction benchmark}

\begin{algorithm}
\caption{Degree-corrected link prediction benchmark}
\label{alg:degree-corrected-benchmark}
\begin{algorithmic}[1]
\State \textbf{Input:} Graph $G(\set{V}, \set{E})$, Sampling fraction $\beta \in [0,1]$ for positive edges
\State \textbf{Output:} Set of negative edges $\set{E}_{\text{neg}}$ and set of positive edges $\set{E}_{\text{pos}}$
\State Generate $\set{E}_{\text{pos}}$ by randomly sampling $\beta$ fraction of edges in $\set{E}$.
\State Initialize $\set{E}_{\text{neg}} \gets \emptyset$
\State Create a node list $L$ where each node $i \in \set{V}$ with degree $k_i$ appears $k_i$ times
\While{$|\set{E}_{\text{neg}}| < |\set{E}_{\text{pos}}|$}
    \State Randomly select two nodes $i, j$ from $L$ with replacement
    \If{$(i, j) \notin \set{E}$ and $(i, j) \notin \set{E}_{\text{neg}}$ and $i \neq j$}
        \State $\set{E}_{\text{neg}} \gets \set{E}_{\text{neg}} \cup \{(i, j)\}$
    \EndIf
\EndWhile
\State \textbf{return} $\set{E}_{\text{neg}}$, $\set{E}_{\text{pos}}$
\end{algorithmic}
\end{algorithm}

The link prediction benchmark yields biased evaluations due to mismatched degree distributions between positive and negative edges, i.e., $p_\text{neg}(k)\neq p_\text{pos}(k)$.
To mitigate the mismatch, we introduce \emph{the degree-corrected link prediction benchmark} that samples negative edges with the same degree bias as positive edges (Algorithm~\ref{alg:degree-corrected-benchmark}).
Specifically, we create a list of nodes where each node with degree $k$ appears $k$ times.
Then, we sample negative edges by uniformly sampling two nodes from this list with replacement until the sampled node pairs are not connected and not in the test edge set.
Crucially, nodes with degree $k$ are $k$ times more likely to be sampled than nodes with degree 1, mirroring the degree bias of the positive edges.
Consequently, the positive and negative edges in the degree-corrected benchmark are indistinguishable based on node degrees.
A Python package for the degree-corrected benchmark is available at \url{https://github.com/skojaku/degree-corrected-link-prediction-benchmark}.

\subsubsection*{Comparison of the benchmark evaluations}

\begin{figure}
    \centering
    \includegraphics[width=\hsize]{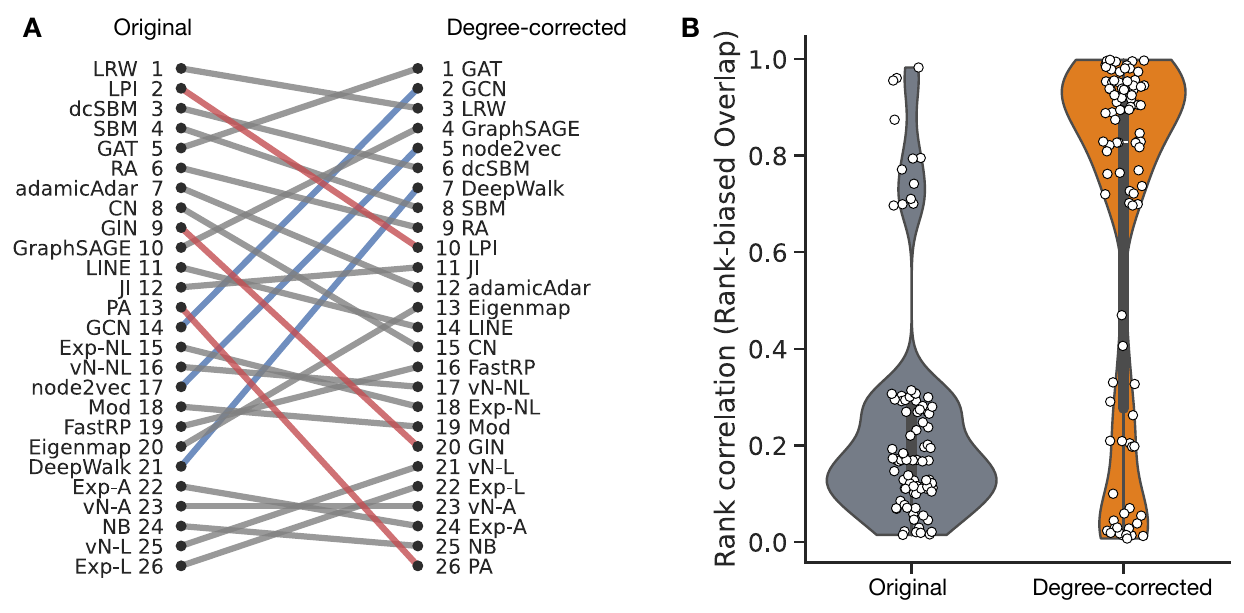}
    \caption{
        Comparative analysis of link prediction and recommendation benchmarks.
        {\bf A}: Rank changes across the original and degree-corrected link prediction benchmarks. The red and blue lines indicate the methods that change their rankings more than 8 places from the original one.
        We compute the ranking of the methods based on the AUC-ROC of the link prediction benchmark and compare it against the
        ranking based on the link retrieval task.
        {\bf B}: The degree-corrected benchmark ranks link prediction methods more similarly with the recommendation task than the original benchmark.
        The RBO (rank-biased overlap) represents the similarity between the ranking of the link prediction methods and that based on the recommendation task.
    }
    \label{fig:uniform-vs-biased}
\end{figure}

We reevaluated the methods with the degree-corrected link prediction benchmark (Fig.~\ref{fig:bias-illustration}E).
We use the same parameters in the original link prediction benchmark for all methods.
There is some agreement between the original and degree-corrected benchmarks (Fig.~\ref{fig:uniform-vs-biased}A).
For example, they rank \method{GAT} and \method{LRW} as top performers, while\method{NB}, \method{SGTAdjNeu}, and \method{SGTAdjExp} are consistently ranked lower.
On the other hand, the degree-corrected benchmark ranks \method{PA} as the lowest performer, with its average AUC-ROC dropping from 0.83 to 0.54, placing it last out of \numMethods methods.
The AUC-ROC near 0.5 for \method{PA} suggests that it performs as random guessing, confirming that the degree-corrected benchmark successfully makes the degree distributions of positive and negative edges indistinguishable.
Other methods such as \method{LPI}, \method{GIN} also experience a substantial drop in their rankings from 2nd to 10th and 9th to 20th, respectively (Fig.~\ref{fig:uniform-vs-biased}A).
On the other hand, \method{GCN}, \method{node2vec}, \method{DeepWalk}, and \method{EigenMap} increase their rankings substantially from 14th to 2nd, 17th to 5th, 21th to 7th, and 20th to 12th, respectively (Fig.~\ref{fig:uniform-vs-biased}A).

\subsubsection*{The degree-corrected benchmark aligns better with recommendation tasks}

Link prediction methods are often used in recommendation tasks~\cite{li2024evaluating,menand2024link,yang2015evaluating,huang2023link,wang2021pairwise}.
However, there are key differences between standard link prediction benchmarks and recommendation tasks~\cite{li2024evaluating,menand2024link,yang2015evaluating}.
In the link prediction benchmark, a set of potential edges is provided, and the goal is to determine the existence of these edges.
In contrast, recommendation tasks require identifying potential edges from the entire graph without a preset list of candidate edges.
This makes recommendation tasks more challenging due to the large number of unconnected nodes that could be just as likely to have an edge as connected nodes.
For example, a link prediction task in a social network involves predicting friendships from a given list of potential friends.
On the other hand, recommendation tasks must evaluate all potential unlisted friendships, including absent but equally probable friendships as existing ones.
Nevertheless, given the common use of link prediction in recommendation tasks, it is crucial that the link prediction benchmarks accurately reflect the performance in the recommendation tasks.

We assess how well the link prediction benchmark provides coherent performance assessment with the recommendation task as follows.
In the common recommendation task~\cite{menand2024link}, the method recommends, for each node $i$, its top $C=50$ nodes $j$ based on the highest scores $s_{ij}$.
We note that our results are consistent regardless of the chosen $C$ value (refer to SI Section~2.3).
The effectiveness of the recommendations is then measured using the vertex-centric max precision recall at $C$ (VCMPR@C) defined as the higher of the precision and recall scores at $C$~\cite{menand2024link}.
The VCMPR@C is designed to evaluate recommendation methods and addresses the excessive penalty on the precision scores at $C$ for small-degree nodes, which often have low precision due to their limited number of edges.
We perform this task five times and average the VCMPR@C scores across different runs.

For each graph, we evaluate the alignment between the rankings based on the recommendation task and those based on the link prediction benchmarks using Rank Biased Overlap (RBO)~\cite{WebberASimilarityMeasure2010}.
RBO is a ranking similarity metric with larger weights on the top performers in the two rankings.
A larger RBO score indicates that the top performers in the two rankings are more similar.
The weights on the top performer are controlled by the parameter $p \in (0,1)$.
While we set $p=0.5$ in our experiment, we confirmed that our results are robust to the choice of $p$ (SI Section~2.3).

Our results from \numNetworks graphs show that the degree-corrected benchmark achieves higher RBO scores than the original benchmark. We find consistent results for different values of $C$ and parameter $p$ of the RBO (SI Section~2.3).
These results indicate that the degree-corrected benchmark more accurately mirrors the performance in recommendation tasks, providing a more reliable measure of the effectiveness of methods in practical applications.

\subsubsection*{Degree-corrected benchmark facilitates the learning of community structure}
\label{sec:training_framework}

\begin{figure}
    \centering
    \includegraphics[width=\hsize]{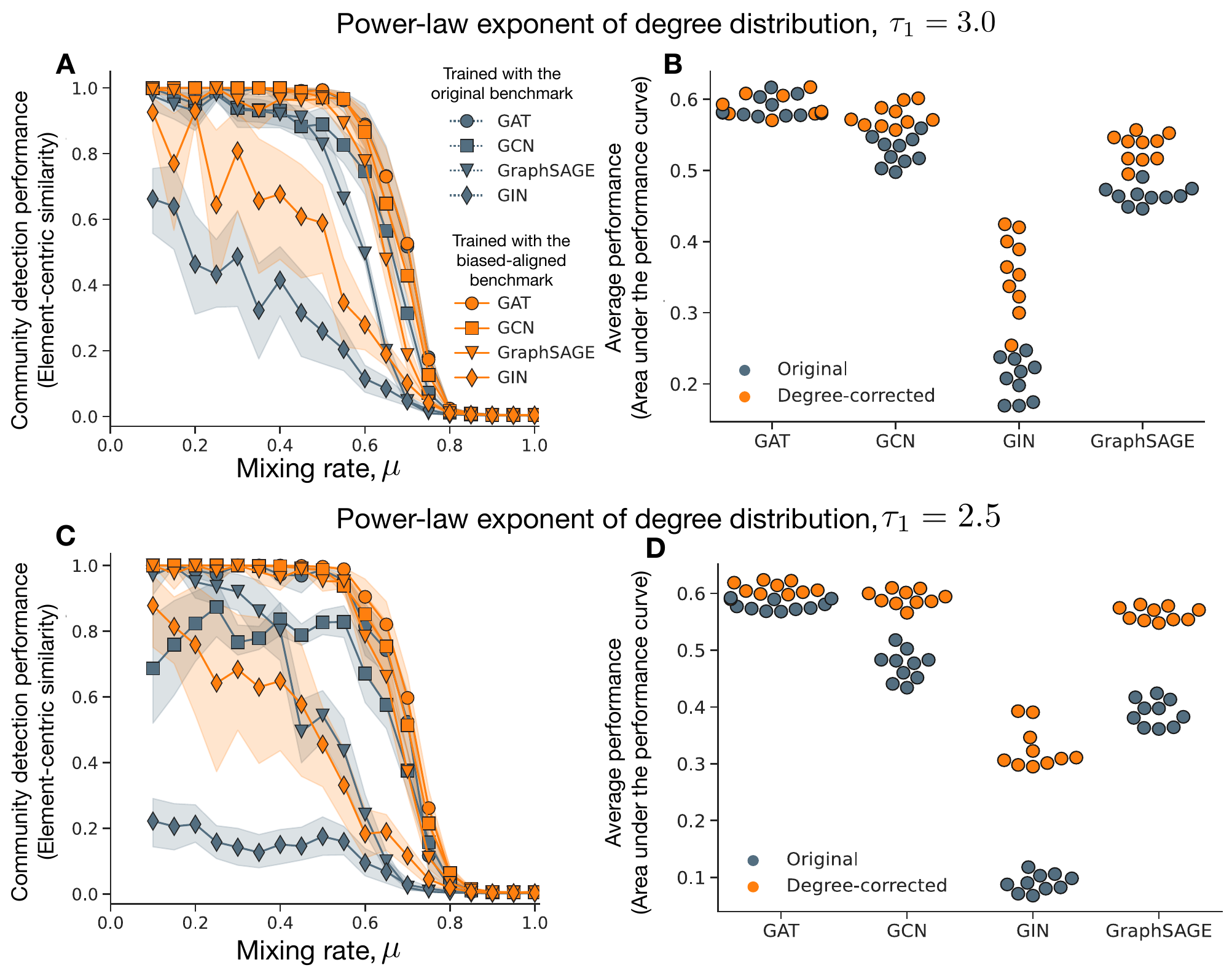}
    \caption{
        The degree-corrected benchmark improves GNNs in learning community structure in the LFR graphs.
        The LFR graphs consist of $3,000$ nodes with the average degree of $25$.
        {\bf A}: The performance of community detection for the LFR graphs with a power-law degree distribution with $\tau_1 = 3.0$ as a function of $\mu$.
        {\bf B}: The average performance (by the area under the performance curve) shows that the methods trained with the degree-corrected training set outperform their counterparts trained with the traditional link prediction training set.
        {\bf C, D}: The same plots for LFR graphs with a fatter power-law degree distribution with $\tau_1 = 2.5$.
        The error bars represent the 95\% confidence interval estimated by a bootstrap of 1,000 repetitions.
    }
    \label{fig:lfr-benchmark}
\end{figure}

The link prediction benchmark is a common unsupervised learning objective for GNNs~\cite{hamilton2017inductive,kawamoto2018mean,kojaku2021residual2vec,tangLINELargescaleInformation2015}.
The degree bias implies that GNNs trained using the original link prediction benchmark tend to overfit to node degrees because they can easily differentiate between positive and negative edges based on node degrees.
We show that the degree-corrected benchmark effectively prevents overfitting to node degrees and improves the learning of salient graph structures.

We evaluate GNNs on the common unsupervised task of community detection in graphs~\cite{fortunato20YearsNetwork2022,fortunatoCommunityDetectionGraphs2010,fortunatoCommunityDetectionNetworks2016}.
The community detection task identifies densely connected groups (i.e., communities) in a graph.
Communities often correspond to functional units (e.g., social circles with similar opinions and protein complexes) in the graph, and detecting communities is a crucial task in many graph applications~\cite{fortunato20YearsNetwork2022,fortunatoCommunityDetectionGraphs2010,fortunatoCommunityDetectionNetworks2016,peixotoParsimoniousModuleInference2013,peixoto2018reconstructing}.
Specifically, we test the GNNs by using the Lancichinetti-Fortunato-Radicchi (LFR) community detection benchmark~\cite{lancichinettiBenchmarkGraphsTesting2008}, a standard benchmark for community detection~\cite{fortunatoCommunityDetectionNetworks2016,fortunatoCommunityDetectionGraphs2010,tandonCommunityDetectionNetworks2021,kojaku2023network}.
The LFR benchmark generates synthetic graphs with predefined communities as follows.
Each node $i$ is assigned a degree $k_i$ from a power-law distribution $p(k) \sim k^{-\tau_1}$, with maximum degree $k_{\text{max}}$.
Nodes are randomly grouped into $L$ communities, with community sizes (i.e., the number of nodes in a community) following another power-law distribution $p(n) \sim n^{-\tau_2}$ bounded between $n_{\text{min}}$ and $n_{\text{max}}$.
Edges are then formed such that each node $i$ connects to a fraction $1-\mu$ of its $k_i$ edges within its community and the remaining fraction $\mu$ to nodes in other communities.
We generate 10 graphs for $\mu \in \{0.05, 0.1, 0.15,\ldots, 0.95, 1\}$ using the following parameter values: the number of nodes $N=3,000$, the degree exponent $\tau_1 \in \{2.5, 3\}$, the average degree $\langle k \rangle=25$, the maximum degree $k_{\text{max}}=1000$, the community-size exponent $\tau_2 =3$, the minimum and maximum community size $n_{\text{min}}=100$ and $n_{\text{max}}=1000$. We obtained qualitatively similar results for different values of the parameters of the LFR benchmark (SI Section~2.5).

We train GNNs using either the original or the degree-corrected link prediction benchmarks to minimize binary entropy loss in classifying the positive and negative edges.
Using the trained GNNs, we generate node embeddings and apply the $K$-means clustering algorithm to detect communities, where $K$ is set to the number of true communities.
Although the number $K$ of communities is often unknown, we use the ground-truth number to eliminate noise from estimating $K$ and to concentrate on evaluating the quality of the learned node embeddings, a standard practice in benchmarking node embeddings for the community detection task~\cite{tandonCommunityDetectionNetworks2021,kojaku2023network,kovacs2024iterative}.
We measure the performance of GNNs by comparing the detected communities against the true communities using the adjusted element-centric similarity~\cite{kojaku2023network,kovacs2024iterative,gatesElementcentricClusteringComparison2019}, where higher scores indicate a higher similarity between the node partitions for the true and detected communities.
We observe qualitatively similar results for partition similarities based on the normalized mutual information (SI Section~2.4).

All GNNs, except for \method{GIN}, perform well when $\mu \leq 0.5$, where communities are distinct and easily identifiable, but their performance declines as $\mu$ increases (Fig.~\ref{fig:lfr-benchmark}A).
Across a broad range of $\mu$, degree-corrected GNNs, particularly \method{GIN}, \method{GCN}, and \method{GraphSAGE}, outperform original GNNs in identifying communities, as shown by the area under the performance curve (Fig.~\ref{fig:lfr-benchmark}B).
The advantage of degree-corrected benchmarks becomes more evident with a more heterogeneous degree distribution (Fig.~\ref{fig:lfr-benchmark}C and D).
This indicates that degree correction effectively reduces overfitting to node degrees, enhancing the learning of community structures in graphs.

\section{Discussion}
\label{sec:discussion}

We showed that the common link prediction benchmark suffered from a sampling bias arising from node degree and could mislead performance evaluations by favoring methods that overfit to node degree.
To address this issue, we proposed a degree-corrected benchmark that corrected the discrepancy between the degree distributions of the connected and unconnected node pairs sampled for performance evaluation.
This new benchmark provided more accurate evaluations that aligned better with the performance for recommendation tasks and also improved training for GNNs by reducing the overfitting to node degree and facilitating the learning of communities in the graph.

While we focused on node degree, other features can also effectively distinguish between positive and negative edges~\cite{li2024evaluating,lichtnwalter2012link,zhang2018link,mao2023revisiting}.
Previous studies showed that nodes in the negative edge set tend to be distant from each other, making them easily identifiable by computing the distance between nodes~\cite{li2024evaluating,lichtnwalter2012link}.
Our findings add a new direction to the important examination of the fundamental graph machine-learning task by demonstrating that node degree, a simpler structural attribute, is sufficient, in many cases, for differentiating the positive and negative edges.
We note that the degree bias partially accounts for the distance discrepancy in the positive and negative edges because a high degree heterogeneity can induce loops of short length in the graph~\cite{bianconi2005loops}.
Crucially, the degree bias stems from the degree distribution rather than the connections between nodes and thus is present in any non-regular graph.
More broadly, the degree bias is manifested through the edge sampling process, a general technique for evaluating and training graph machine learning.
For instance, mini-batch training~\cite{hamilton2017inductive,hu2020heterogeneous}, which samples subsets of edges for efficient GNN training, may also exhibit bias due to node degrees, leading to skewed training sets.
Given the widespread use of edge sampling across various graph machine-learning tasks, our findings have broad implications beyond link prediction benchmarks, extending to a range of benchmarks and training frameworks.

Our study has several limitations.
First, we employed standard hyperparameters (e.g., embedding dimensions, GNN layers) across all link prediction methods, achieving reasonable performance in link prediction and community detection tasks. Given the extensive variety of methods and hyperparameters, we did not optimize hyperparameters for specific tasks. Although fine-tuning could potentially enhance performance~\cite{li2024evaluating}, our primary goal was to evaluate the effectiveness of link prediction benchmarks, not to compare the difference in performances between different link prediction methods.
Second, we did not explore the reasons behind the varying performance of different link prediction methods.
It is important to note that there is no single link prediction method that is universally effective for all graphs because the performance of link prediction methods depends on the assumptions on the graph structure to make the predictions~\cite{ghasemian2020stacking}.
Future studies will explore an in-depth examination of the link prediction methods on specific graphs.
Third, we focus on community structure to test the effectiveness of the proposed benchmark as a training framework.
However, other non-trivial graph structures, such as centrality, could be tested through network dismantling benchmarks~\cite{osat2023embedding}.

Even with the aforementioned limitations, our results suggest that sampling graph data is a highly non-trivial task than commonly considered.
Because sampling edges from a graph is integral to evaluating and training graph machine learning methods, our results underline the importance of careful sampling to ensure the effectiveness of evaluations and training of graph machine learning methods.

\section{Broader Impacts}
\label{sec:broader_impact}

Our findings indicate that the standard benchmark promotes link prediction algorithms that enhance the ``rich-get-richer'' effect, where well-connected nodes tend to gain even more connections. This effect, known as \emph{preferential attachment}, is common in various societal contexts.
For instance, popular individuals often gain more friends, widely shared content spreads further, and frequently cited papers are more likely to attract additional citations.
While preferential attachment can benefit distributing resources and recognition, it also presents significant disadvantages.
For example, preferential attachment can disproportionately favor well-connected individuals in professional networks and hinder scientific progress by undervaluing innovative papers with less citations~\cite{chu2021slowed}.
Our analysis indicates that the recommendation algorithms trained using the existing benchmark may exacerbate social inequalities by reinforcing the preferential attachment mechanism.
Our new benchmark directly addresses these concerns by negating the degree bias, promoting fairer algorithm evaluations, and supporting the development of fairness-aware machine learning methods.

\subsubsection*{Code and Data Availability}

We provide the code to reproduce the results in this study at \url{https://github.com/skojaku/degree-corrected-link-prediction-benchmark}.
The network data used in this study is available at \url{https://figshare.com/projects/Implicit_degree_bias_in_the_link_prediction_task/205432}.

\begin{ack}
R.A., M.K., O.C.S., S.K., and Y.Y.A. acknowledge the support by the Air Force Office of Scientific Research under award number FA9550-19-1-0391.
The authors acknowledge the computing resources at Binghamton University and Indiana University and thank NVIDIA Corporation for their GPU resources.
\end{ack}

{
\small


}



\section*{Supplementary material}

# Supplementary Information for "Implicit degree bias in the link prediction task"

# Contents

# 1 Data and link prediction methods

## 1.1 Network data

The corpus of networks used in this work comprises networks with the number of nodes in the range $[10^2, 10^6]$ and edges in the range $[10^2, 10^8]$. This includes social, technological, information, biological, and transportation (spatial) networks. For simplicity in our analysis, we consider these networks to be unweighted, undirected, and without self-loops, though the message of our work holds without these constraints. The largest networks in our corpus (number of nodes $> 10^5$) are sourced from Netzschleuder [1], and the remaining networks are obtained from the authors of Ref. [2]. See Table 2 for details.

## 1.2 Link prediction algorithms

We use 26 link prediction algorithms categorized into four groups: topology-based, graph embedding, network model, and graph neural networks (see Table 1).

### 1.2.1 Topology-based predictors

Topology-based predictors calculate the prediction score $s_{ij}$ using the structural features of two nodes. The topology-based predictors employed in our study include Preferential Attachment (PA) [3], Common Neighbors (CN) [4], Adamic-Adar (AA) [5], Jaccard Index (JI) [4], Resource Allocation (RA) [6, 7], Local Random Walk (LRW) [8], Local Path Index (LPI) [9]. For LRW and LPI, we set the hyperparameter $\epsilon = 0.001$ as per previous studies [8, 9]. The other methods do not require hyperparameters.

### 1.2.2 Graph embeddings

Graph embedding maps a graph into a vector space, with each node $i$ represented by a point in this space. The prediction score $s_{ij}$ is given by the dot product $\boldsymbol{u}_i^\top \boldsymbol{u}_j$ between any two node vectors. We tested a variety of graph embedding methods including Laplacian EigenMap (EigenMap) [10], Spectral Modularity (Mod) [11], Non-backtracking Embedding (NB) [12], FastRP (FastRP) [13], Exponential Kernel on Adjacency Matrix (Exp-A) [14, 15], Exponential Kernel on Laplacian (Exp-L) [14, 15], Exponential Kernel on

Normalized Laplacian (`Exp-NL`) [14, 15], Von Neumann Kernel on Adjacency Matrix (`vN-A`) [15, 16], Von Neumann Kernel on Laplacian (`vN-L`) [15, 16], and Von Neumann Kernel on Normalized Laplacian (`vN-NL`) [15, 16], node2vec (`node2vec`) [17], DeepWalk (`DeepWalk`) [18], and LINE (`LINE`) [19]. We tested a variety of graph embedding methods including Laplacian EigenMap (`EigenMap`) [10], Spectral Modularity (`Mod`) [11], Non-backtracking Embedding (`NB`) [12], FastRP (`FastRP`) [13], Exponential Kernel on Adjacency Matrix (`Exp-A`) [14, 15], Exponential Kernel on Laplacian (`Exp-L`) [14, 15], Exponential Kernel on Normalized Laplacian (`Exp-NL`) [14, 15], Von Neumann Kernel on Adjacency Matrix (`vN-A`) [15, 16], Von Neumann Kernel on Laplacian (`vN-L`) [15, 16], and Von Neumann Kernel on Normalized Laplacian (`vN-NL`) [15, 16], node2vec (`node2vec`) [17], DeepWalk (`DeepWalk`) [18], and LINE (`LINE`) [19]. For all methods, we set the number of embedding dimensions to 128. For `LINE`, `node2vec`, and `DeepWalk`, we set the number of walkers to 40 and the number of the walk length to 80 following Ref. [20]. We used the default hyperparameters used in the original papers unless otherwise specified.

### 1.2.3 Graph Neural Networks

Graph neural networks (GNNs) learn the vector representation, $\boldsymbol{u}_i$, for each node $i$ of the network by using neural networks. The prediction score $s_{ij}$ is given by the dot product $\boldsymbol{u}_i^\top \boldsymbol{u}_j$ between any two node vectors. We also explore several graph neural network (GNN) architectures for link prediction, leveraging the PyTorch Geometric library [21]. The GNN methods we employ include: Graph Convolutional Network (`GCN`) [22], Graph SAGE (`GraphSAGE`) [23], Graph Attention Network (`GAT`) [24], and Graph Isomorphism Network (`GIN`) [25]. We use two hidden layers of 256 dimensions with ReLu activation and a linear output layer of 128 dimensions. The node features are the 64 principal eigenvectors of the adjacency matrix, and we extend the feature vector by adding a 64-dimensional vector with each element being generated from an independent Gaussian distribution with mean 0 and standard deviation 1 by following Ref. [26, 27]. We train GNNs on the link prediction task for 250 epochs with a dropout rate of 0.2, using the Adam optimizer at a learning rate 0.01. We use the 'LinkNeighborLoader' from PyTorch Geometric to generate training mini-batches. This loader samples both positive and negative edges, along with 30 immediate neighbors and 10 secondary neighbors sampled by random walks for each node involved in

Table 1: Link prediction algorithms. "pyg" refers to PyTorch Geometric.

| | Algorithm | Reference | Code | Notation |
|---|---|---|---|---|
| Topology based | Preferential attachment | [3] | ourselves | PA |
| | Common neighbors | [4] | ourselves | CN |
| | AdamicAdar | [5] | ourselves | AA |
| | Jaccard index | [4] | ourselves | JI |
| | Resource allocation | [6, 7] | ourselves | RA |
| | Local Random Walk | [8] | ourselves | LRW |
| | Local Path Index | [9] | ourselves | LPI |
| Graph embedding | Laplacian EigenMap | [10] | ourselves | EigenMap |
| | Spectral modularity | [11] | ourselves | Mod |
| | Non-backtracking embedding | [12] | ourselves | NB |
| | FastRP | [13] | ourselves | FastRP |
| | Adjacency matrix w/ the exponential kernel | [14, 15] | ourselves | Exp-A |
| | Laplacian w/ the exponential kernel | [14, 15] | ourselves | Exp-L |
| | Normalized Laplacian w/ the exponential kernel | [14, 15] | ourselves | Exp-NL |
| | Adjacency matrix w/ the von Neumann kernel | [15, 16] | ourselves | vN-A |
| | Laplacian w/ the von Neumann kernel | [15, 16] | ourselves | vN-L |
| | Normalized Laplacian w/ the von Neumann kernel | [15, 16] | ourselves | vN-NL |
| | LINE | [19] | gensim [34, 35] | LINE |
| | DeepWalk | [18] | gensim [34, 35] | DeepWalk |
| | node2vec | [17] | gensim [34, 35] | node2vec |
| | node2vec | [17] | gensim [34, 35] | node2vec |
| Graph neural networks | Graph Convolutional Network | [22] | pyg [21] | GCN |
| | Graph SAGE | [23] | pyg [21] | GraphSAGE |
| | Graph SAGE | [23] | pyg [21] | GraphSAGE |
| | Graph Attention Network | [24] | pyg [21] | GAT |
| | GIN | [25] | pyg [21] | GIN |
| Network model | Stochastic block model | [28–31] | graph tool [33] | SBM |
| | Degree-corrected stochastic block model | [31–33] | graph tool [33] | dcSBM |

these edges [23]. The batch size is set to 5000.

### 1.2.4 Nework models

We use two stochastic block models (SBM) [28–31] and the degree-corrected SBM [31–33]. These models estimate the probability $P(i,j)$ that an edge exists between two nodes, which serves as the prediction score $s_{ij}$. We fit the SBMs using the graph tool package [33]. We select the number of blocks that minimize the description length and use default settings for other parameters.

## 1.3 Additional analysis methods

We use the following additional analysis methods in this paper. We fit a log-normal distribution to the degree distribution of the graphs by using the moment method implemented in the `scipy.stats.lognorm` package [36]. We fit a power-law distribution to the degree distribution of the graphs by using

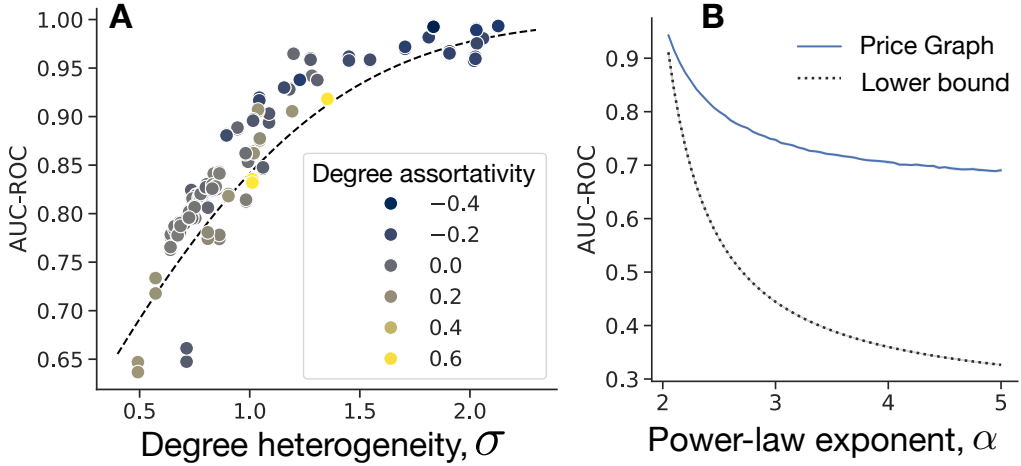

Figure 1: The AUC-ROC for `PA` as a function of degree heterogeneity. **A**: The AUC-ROC for the empirical graphs and that expected by node degree (Eq. (9) in the main text). The colors represent the degree assortativity. **B**: Lower bound for the AUC-ROC for the power-law degree distributions. The dashed line represents the lower bound for the AUC-ROC for the power-law distribution. The blue line represents the AUC-ROC for `PA` for the Price graph with $N = 10^4$ nodes and $M = 10^5$ edges.

the maximum likelihood method implemented in the `powerlaw` package [37]. We compute the RBO score by using the `rbo` package [38].

## 2 Robustness analysis

### 2.1 Impact of degree assortativity on the AUC-ROC for `PA`

We have assumed that the graph has no degree assortativity, meaning that $P(k_i, k_j) = P(k_i)P(k_j)$. Although this assumption may not always hold, it provides a good approximation for the AUC-ROC behavior for `PA`. Although the assortativity varies across graphs, the AUC-ROC for `PA` still closely follows Eq. (9) in the main text (Fig. 1).

5

## 2.2 AUC-ROC for `PA` for scale-free networks

We have assumed that the graph exhibits the heterogeneous degree distributions characterized by the log-normal distribution. An alternative model of the degree distribution is the power-law distribution [39]. Here, we show that our results also hold for the power-law degree distribution, i.e., the AUC-ROC for `PA` increases as the degree heterogeneity increases.

We compute the AUC-ROC for `PA` for graphs with power-law degree distribution. Computing AUC-ROC $P(k_{i-}k_{j-} \leq k_{i+}k_{j+})$ is not trivial because it involves multiplicative convolution of two probability distributions, which are hard to compute for the power law degree distribution. To circumvent this problem, we consider the lower-bound by focusing on $k_{i-} \leq k_{i+}$ and $k_{j-} \leq k_{j+}$, which is the subset of all combinations of $(k_{i-}, k_{i+}, k_{j-}, k_{j+})$ leading to $k_{i-}k_{j-} \leq k_{i+}k_{j+}$, i.e.,

$$P(k_{i-} < k_{i+}) \cdot P(k_{j-} < k_{j+} \mid k_{i-}, k_{i+}) \leq P(k_{i-}k_{j-} < k_{i+}k_{j+}) \quad (1)$$

Assuming that the graph has no degree assortativity (i.e., $P(k_i, k_j) = P(k_i)P(k_j)$), we obtain the lower bound for the AUC-ROC:

$$P(k_{i-} < k_{i+}) \geq P(k_{i-} < k_{i+})^2 = \left[\sum_{k=1}^{\infty} p_{\text{neg}}(k) \sum_{\ell=k}^{\infty} p_{\text{pos}}(\ell)\right]^2. \quad (2)$$

Now, let us compute the lower bound by assuming that the degree distribution follows a power-law [40]:

$$p(k) = \frac{1}{\zeta(\alpha, k_{\min})} k^{-\alpha}, \quad (k \geq k_{\min}), \text{ where } \zeta(\alpha, k_{\min}) = \sum_{\ell=k_{\min}}^{\infty} \ell^{-\alpha}, \quad (3)$$

where $\zeta$ is the Hurwitz zeta function, and $k_{\min}$ is the minimum degree. By substituting Eq. (1) in the main text into Eq. (3), we have $p_{\text{pos}} = k^{-\alpha+1}/\zeta(\alpha-1, k_{\min})$. By noting that $\sum_{\ell=k}^{\infty} p(\ell) = \zeta(\alpha, k)/\zeta(\alpha, k_{\min})$ [40], we have

$$P(k_{i-} < k_{i+})^2 = \left[\frac{1}{\zeta(\alpha, k_{\min})\zeta(\alpha-1, k_{\min})} \sum_{k=k_{\min}}^{\infty} k^{-\alpha} \zeta(\alpha-1, k)\right]^2. \quad (4)$$

Numerical calculation shows that the lower bound $P(k_{i-} < k_{i+})^2$ approaches 1 as $\alpha \to 2$ (Fig. 1). Additional validation using the Price network with

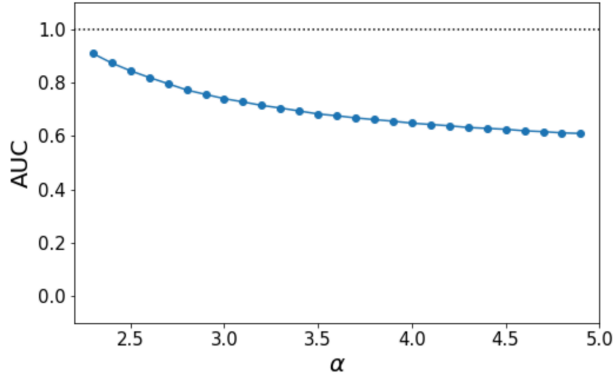

Figure 2: The influence of degree heterogeneity on the performance of the preferential attachment (PA) link prediction model in graphs with a power law degree distribution. The degree heterogeneity is governed by the power law exponent $\alpha$. As $\alpha$ increases, the heterogeneity decreases. PA reaches near maximal AUC-ROC scores (1) as $\alpha \to 2$. For each $\alpha$, we generate 20 batches, each with 5000 samples of $k_i^+, k_j^+, k_m^-, k_j^-$. The dots indicate the average AUC score obtained via the Mann-Whitney U statistic. Standard mean errors are smaller than the dots.

$N = 10^4$ nodes and $M = 10^5$ edges, where $p(k) \propto k^{-\alpha}$, confirms that PA achieves higher AUC-ROC than the lower-bound and reaches near-maximal AUC-ROC scores for $\alpha \approx 2$.

We can also compute the AUC score using the Mann-Whitney U statistic (Fig. 2). Let us take a graph $G$ with the set of nodes given by $\mathcal{V}$. We sample nodes $i, j$ with degrees $k_i^+, k_j^+$ forming the positive set of edges from $p_{pos}$, and nodes $m, n$ with degrees $k_m^-, k_n^-$ forming the negative set of edges from $p_{neg}$. Then $P(k_i^+ k_j^+ > k_m^-, k_n^-) \ \forall i, j, k, m \in \mathcal{V}$ is the AUC score and is given by $\frac{U}{n_1 n_2}$ where U is the Mann-Whitney U statistic and $n_1, n_2$ are sizes of the positive and negative edge sets respectively [41]. We sample the random variables $k_i^+, k_j^+$ using the "Power_Law" function from the powerlaw package [37] with degree exponent $\alpha - 1$ since $p_{pos}(k) \sim k^{-(\alpha-1)}$. Similarly, we sample $k_m^-, k_n^-$ with degree exponent $\alpha$ since $p_{neg}(k) \sim k^{-\alpha}$. Fig. 2 aligns with our findings in Fig. 1B, i.e., PA reaches near maximal AUC-ROC scores as $\alpha \to 2$.

7

## 2.3 Parameter sensitivity in the analysis of performance alignment with the recommendation task

The vertex-centric max precision recall at $C$ (VCMPR@C) metric [42] is a metric computed based on the precision and recall of the recommendations for each node. This metric is proposed for evaluating link prediction methods in recommendation settings. The VCMPR@C for a node $i$ and recommended node set $\mathcal{V}_i$ is defined as

$$\text{VCMPR@C for node } i = \frac{\sum_{j \in \mathcal{V}_i} Y_{ij}}{\max(C, m_i)}, \tag{5}$$

where $Y_{ij}$ is the indicator function of node $j$ is connected with $i$ in the test data ($Y_{ij} = 1$), and otherwise $Y_{ij} = 0$. Variable $m_i$ is the number of true connections in the test data, i.e., $m_i = \sum_j Y_{ij}$. We compute the average VCMPR@C for all nodes as the performance of the link prediction method for the graph.

We compute the similarity of two rankings with rank-biased overlap (RBO) [43]. RBO assesses the similarity of two rankings by examining the overlap of top-performing methods. Define $U_{k,1}$ as the set of methods ranked in the top $k$ positions in ranking 1, and $U_{k,2}$ similarly for ranking 2. Then, RBO computes a weighted average of the similarity of the top $k$ methods by

$$\text{RBO}(S, T, p) := (1-p) \sum_{k=1}^{\infty} p^{k-1} \frac{|U_{k,1} \cap U_{k,2}|}{k}, \tag{6}$$

where $p$ controls the importance of the top performer, with a smaller $p$ value placing more weight on the top performer. We use $p = 0.5$ for the results in the main text. We find consistent results across different $p$ values (Fig. 3A and B). Additionally, we find consistent results for a different number of recommendations $C$ (Fig. 3C and D).

## 2.4 Evaluation of the community detection performance using the normalized mutual information

Normalized Mutual Information (NMI) is a standard metric for assessing community detection methods [29, 44]. NMI quantifies the similarity between actual and predicted community assignments, where a score of zero indicates no similarity. We note that NMI has a bias favoring partitions with small

8

communities [45], and thus, we used the element-centric similarity that does not have this bias in our main experiment. We note that NMI has a bias favoring partitions with small communities [45], and thus, we used the element-centric similarity that does not have this bias in our main experiment. Nevertheless, we include the results for NMI in Fig. 4 for comparison. As with the element-centric similarity, our results show that the degree-corrected GNNs perform on par or better than the original GNNs.

## 2.5 Sensitivity to the choice of the LFR benchmark parameters

We tested the robustness of the results by using different parameter values for the LFR benchmark. First, we confirmed the consistent results when varying the average degree $\langle k \rangle$ from 25 to 50 (Fig. 5), or the maximum community size and degree from 1000 to 500 (Fig. 6).

# 3 Reproducibility

### 3.0.1 Source data and code

The source data, code, and workflow for our experiments are available on GitHub and FigShare. The URLs are omitted in accordance with NeurIPS anonymity guidelines; however, we provide the data and code in the supplementary materials.

### 3.0.2 Snakemake workflow

We ensure the reproducibility of our experiments by using Snakemake [46], which allows automatic workflow execution from the preprocessing to the generation of the plots. With the Snakemake workflow, the user can reproduce all results by running the following command in the terminal:

```
snakemake --cores <number of cores> all
```

The workflow requires Python 3.11 or later, and all required Python packages are listed in the "environment.yaml" file in the repository.

### 3.0.3 Execution time and hardware requirements

We run the workflow on a server with 64 Intel(R) Xeon(R) Gold 5218 CPUs equipped with 64 cores, 1T RAM, and four NVIDIA GPUs with 48 GB memory, sufficient to complete the workflow in one week. The execution time of the workflow for the community detection task is 4 days, and that for the link prediction task is 10 days. The workflow can be executed with fewer resources by reducing the number of cores. The minimum computer requirements to run the workflow are as follows:

- 64 GB RAM
- 16 GB GPU memory
- 8 core CPU
- 300 GB space

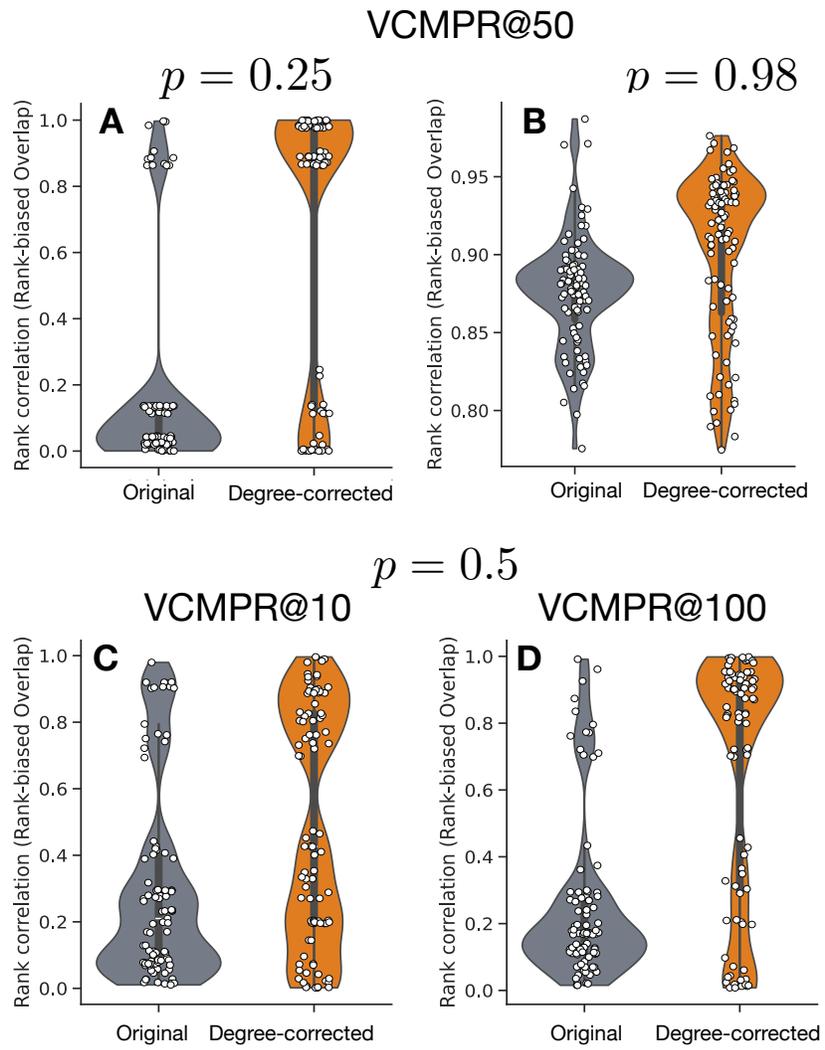

Figure 3: RBO for different $p$ values and different numbers $C$ of recommendations.

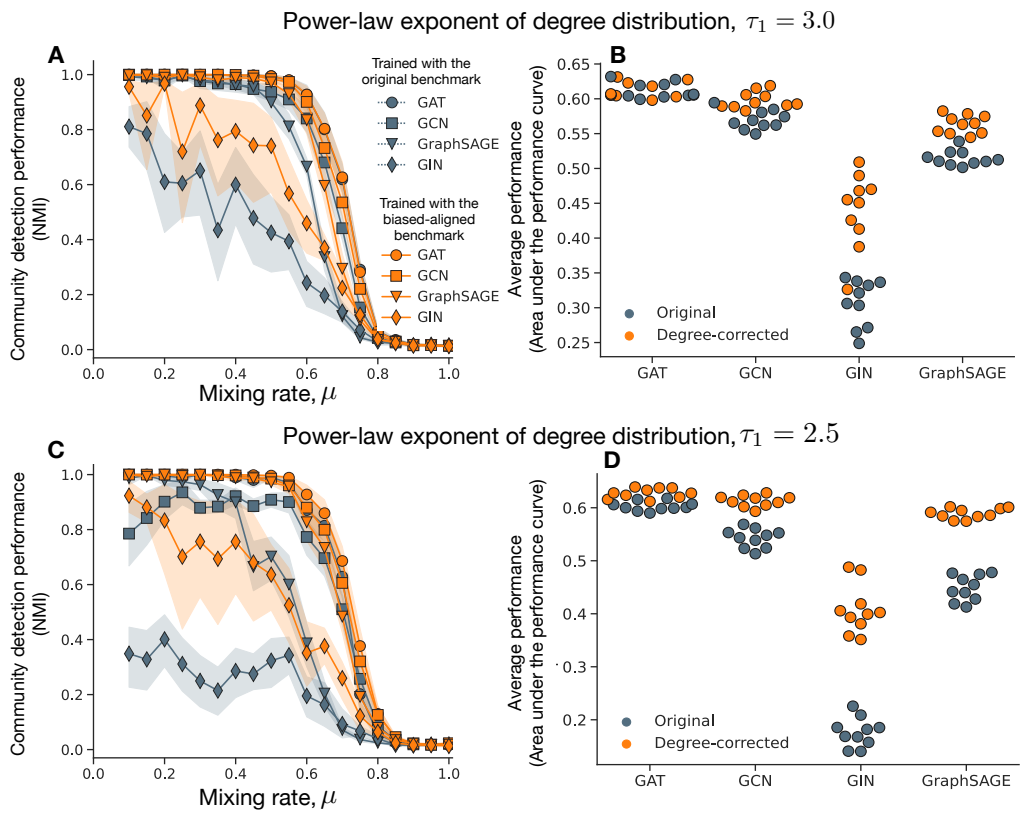

Figure 4: Performance of the GNNs on the LFR benchmark measured by NMI.

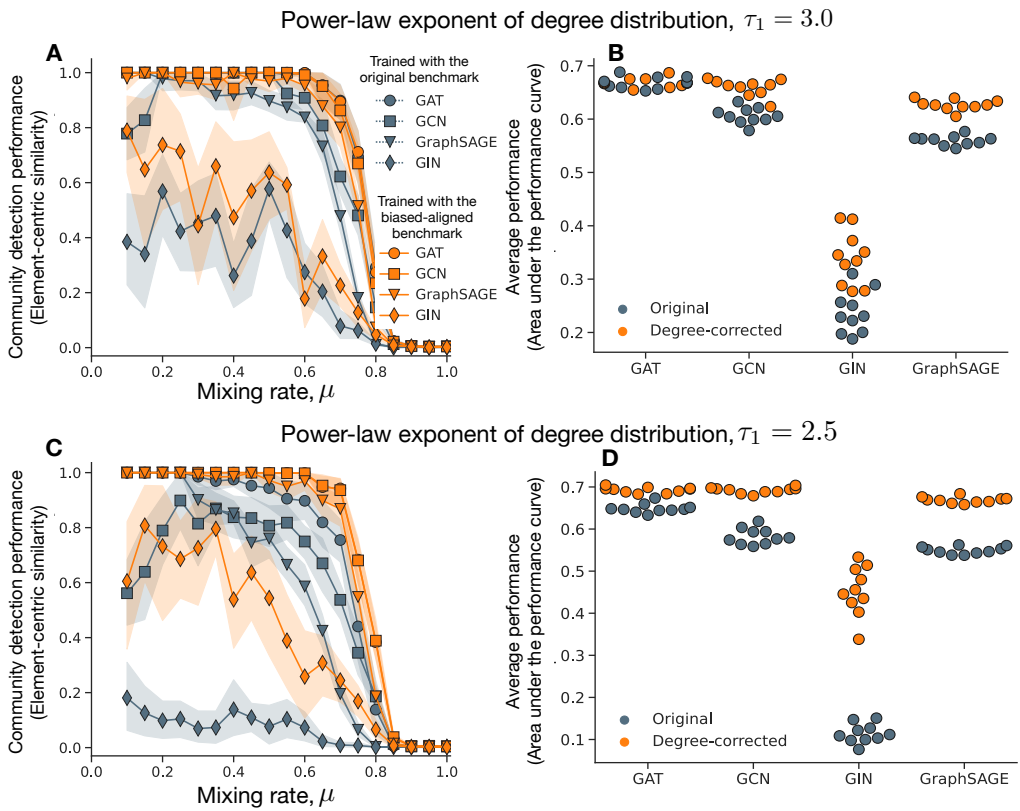

Figure 5: Performance of the GNNs on the LFR benchmark measured by NMI when varying the average degree $\langle k \rangle$ from 25 to 50.

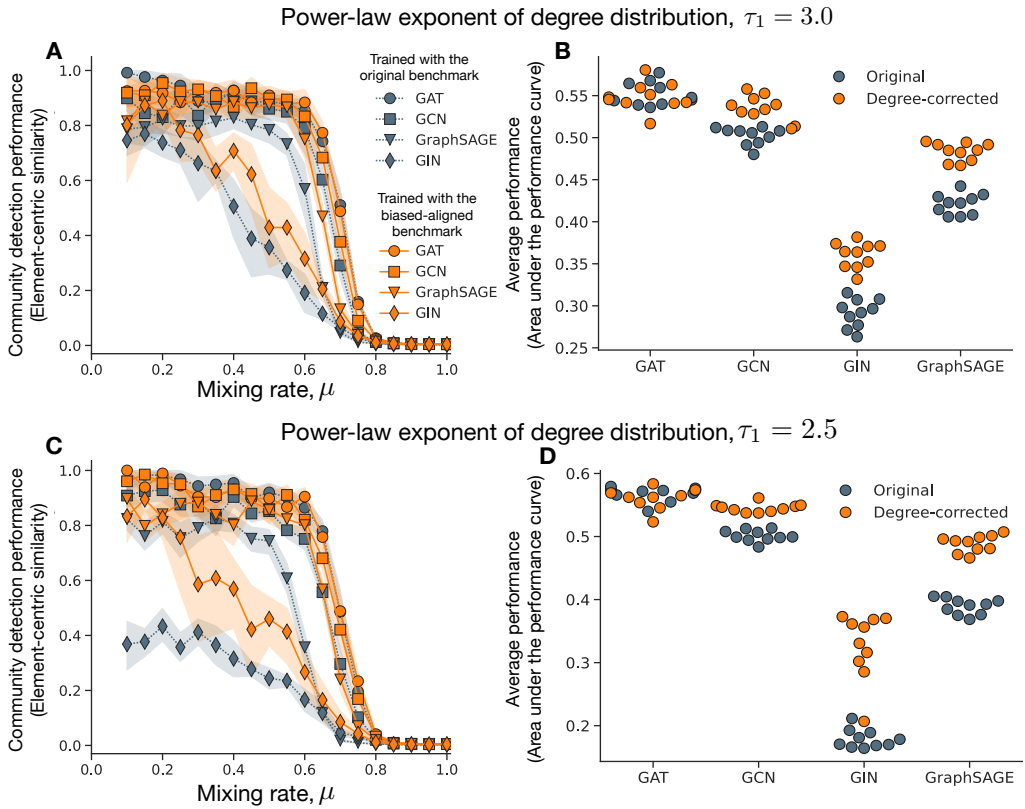

Figure 6: Performance of the GNNs on the LFR benchmark measured by NMI when varying the maximum community size and degree from 1000 to 500.

Table 2: Network data tested in this study. We consider social, technological, information, biological, and transportation (spatial) networks. For simplicity in our analysis, we consider these networks to be unweighted, undirected, and without self-loops. Variance referes to the variance of the node degrees. Assortativity refers to the degree assortativity, and Heterogeneity refers to the degree heterogeneity computed by [47].

| Network | Nodes | Edges | Max. Degree | Variance | Assortativity | Heterogeneity |
|---|---|---|---|---|---|---|
| Political books | 105 | 441 | 25 | 29.69 | -0.128 | 0.43 |
| College football | 115 | 613 | 12 | 0.78 | 0.162 | 0.20 |
| High school 2011 | 126 | 1709 | 55 | 153.71 | 0.083 | 0.62 |
| Food web bay wet | 128 | 2075 | 110 | 249.17 | -0.112 | 0.62 |
| Food web bay dry | 128 | 2106 | 110 | 249.85 | -0.104 | 0.63 |
| Radoslaw email | 167 | 3250 | 139 | 993.84 | -0.295 | 0.62 |
| Highschool 2012 | 180 | 2220 | 56 | 120.37 | 0.046 | 0.51 |
| Little Rock Lake | 183 | 2434 | 105 | 433.29 | -0.266 | 0.54 |
| Jazz | 198 | 2742 | 100 | 303.12 | 0.020 | 0.55 |
| C. Elegans | 297 | 2148 | 134 | 167.56 | -0.163 | 0.38 |
| Network science | 379 | 914 | 34 | 15.42 | -0.082 | 0.23 |
| Dublin social | 410 | 2765 | 50 | 70.51 | 0.226 | 0.29 |
| Airport | 500 | 2980 | 145 | 499.03 | -0.268 | 0.35 |
| Caltech | 762 | 16651 | 248 | 1365.76 | -0.066 | 0.42 |
| Reed | 962 | 18812 | 313 | 1254.53 | 0.023 | 0.38 |
| Political blogs | 1222 | 16714 | 351 | 1474.67 | -0.221 | 0.34 |
| Haverford | 1446 | 59589 | 375 | 3687.70 | 0.067 | 0.41 |
| Simmons | 1510 | 32984 | 300 | 1288.53 | -0.062 | 0.32 |
| Swarthmore | 1657 | 61049 | 577 | 3472.20 | 0.061 | 0.37 |
| Petster | 1788 | 12476 | 272 | 440.86 | -0.089 | 0.24 |
| UC Irvine | 1893 | 13835 | 255 | 599.57 | -0.188 | 0.24 |
| Yeast | 2224 | 6609 | 64 | 63.67 | -0.105 | 0.15 |
| Amherst | 2235 | 90954 | 467 | 4007.71 | 0.058 | 0.35 |
| Bowdoin | 2250 | 84386 | 670 | 3206.35 | 0.056 | 0.33 |
| Hamilton | 2312 | 96393 | 602 | 3940.69 | 0.031 | 0.34 |
| Adolescent health | 2539 | 10455 | 27 | 18.59 | 0.251 | 0.10 |
| Trinity | 2613 | 111996 | 404 | 3742.49 | 0.072 | 0.32 |
| USFCA | 2672 | 65244 | 405 | 2041.31 | 0.092 | 0.27 |
| Japanese book | 2698 | 7995 | 725 | 608.58 | -0.259 | 0.17 |
| Williams | 2788 | 112985 | 610 | 3901.94 | 0.040 | 0.32 |
| Open flights | 2905 | 15645 | 242 | 485.44 | 0.049 | 0.21 |
| Oberlin | 2920 | 89912 | 478 | 2838.11 | 0.050 | 0.28 |
| Wellesley | 2970 | 94899 | 746 | 3079.68 | 0.064 | 0.29 |
| Smith | 2970 | 97133 | 349 | 2432.51 | 0.044 | 0.28 |
| Vassar | 3068 | 119161 | 482 | 3453.23 | 0.101 | 0.30 |
| Middlebury | 3069 | 124607 | 473 | 3865.24 | 0.078 | 0.30 |

Table 3: (Continued) Network data tested in this study.

| Network | Nodes | Edges | Max. Degree | Variance | Assortativity | Heterogeneity |
|---|---|---|---|---|---|---|
| Pepperdine | 3440 | 152003 | 674 | 5695.91 | 0.055 | 0.31 |
| Colgate | 3482 | 155043 | 773 | 4009.14 | 0.067 | 0.29 |
| Santa | 3578 | 151747 | 1129 | 4933.35 | 0.071 | 0.29 |
| Wesleyan | 3591 | 138034 | 549 | 3548.92 | 0.095 | 0.28 |
| Mich | 3745 | 81901 | 419 | 1997.51 | 0.142 | 0.24 |
| Bitcoin alpha | 3775 | 14120 | 511 | 402.89 | -0.169 | 0.17 |
| Bucknell | 3824 | 158863 | 506 | 3498.69 | 0.094 | 0.27 |
| Brandeis | 3887 | 137561 | 1972 | 4646.98 | -0.026 | 0.27 |
| Howard | 4047 | 204850 | 1215 | 8506.41 | 0.058 | 0.32 |
| Rice | 4083 | 184826 | 581 | 5669.22 | 0.065 | 0.29 |
| GR-QC 1993-2003 | 4158 | 13422 | 81 | 74.41 | 0.639 | 0.12 |
| Tennis | 4338 | 81865 | 451 | 4573.31 | 0.003 | 0.26 |
| Rochester | 4561 | 161403 | 1224 | 3632.47 | 0.025 | 0.25 |
| Lehigh | 5073 | 198346 | 973 | 4073.33 | 0.035 | 0.24 |
| JohnsHopkins | 5157 | 186572 | 886 | 4761.94 | 0.080 | 0.25 |
| HT09 | 5352 | 18481 | 1287 | 1333.44 | -0.431 | 0.14 |
| Wake | 5366 | 279186 | 1341 | 7469.92 | 0.071 | 0.27 |
| Hep-Th 1995-99 | 5835 | 13815 | 50 | 20.77 | 0.185 | 0.08 |
| Bitcoin OTC | 5875 | 21489 | 795 | 531.22 | -0.165 | 0.15 |
| Reactome | 5973 | 145778 | 855 | 4612.48 | 0.241 | 0.21 |
| Jung | 6120 | 50290 | 5655 | 16029.25 | -0.233 | 0.16 |
| Gnutella Aug 08 2002 | 6299 | 20776 | 97 | 72.95 | 0.036 | 0.11 |
| American | 6370 | 217654 | 930 | 3847.11 | 0.066 | 0.22 |
| MIT | 6402 | 251230 | 708 | 6241.81 | 0.120 | 0.24 |
| JDK | 6434 | 53658 | 5923 | 16112.86 | -0.223 | 0.16 |
| William | 6472 | 266378 | 1124 | 5164.22 | 0.052 | 0.23 |
| U Chicago | 6561 | 208088 | 1624 | 4093.91 | 0.018 | 0.22 |
| Princeton | 6575 | 293307 | 628 | 6164.10 | 0.091 | 0.24 |
| Carnegie | 6621 | 249959 | 840 | 5674.47 | 0.122 | 0.24 |
| Tufts | 6672 | 249722 | 827 | 4525.50 | 0.118 | 0.22 |
| UC | 6810 | 155320 | 660 | 2297.32 | 0.125 | 0.19 |
| Wikipedia elections | 7066 | 100736 | 1065 | 3332.59 | -0.083 | 0.21 |
| English book | 7377 | 44205 | 2568 | 3699.80 | -0.237 | 0.16 |
| Gnutella Aug 09 2002 | 8104 | 26008 | 102 | 66.74 | 0.033 | 0.09 |

Table 4: (Continued) Network data tested in this study.

| Network | Nodes | Edges | Max. Degree | Variance | Assortativity | Heterogeneity |
|---|---|---|---|---|---|---|
| French book | 8308 | 23832 | 1891 | 1217.86 | -0.233 | 0.12 |
| Hep-Th 1993-2003 | 8638 | 24806 | 65 | 41.61 | 0.239 | 0.08 |
| Gnutella Aug 06 2002 | 8717 | 31525 | 115 | 51.87 | 0.052 | 0.09 |
| Gnutella Aug 05 2002 | 8842 | 31837 | 88 | 54.66 | 0.015 | 0.09 |
| PGP | 10680 | 24316 | 205 | 65.24 | 0.238 | 0.09 |
| Gnutella Aug 04 2002 | 10876 | 39994 | 103 | 48.65 | -0.013 | 0.08 |
| Hep-Ph 1993-2003 | 11204 | 117619 | 491 | 2307.04 | 0.630 | 0.16 |
| Spanish book 1 | 11558 | 43050 | 2986 | 3353.23 | -0.282 | 0.12 |
| DBLP citations | 12495 | 49563 | 709 | 284.34 | -0.046 | 0.10 |
| Spanish book 2 | 12643 | 55019 | 5169 | 6953.72 | -0.290 | 0.11 |
| Cond-Mat 1995-99 | 13861 | 44619 | 107 | 45.70 | 0.157 | 0.07 |
| Astrophysics 1 | 14845 | 119652 | 360 | 472.92 | 0.228 | 0.11 |
| Astrophysics 2 | 17903 | 196972 | 504 | 961.58 | 0.201 | 0.11 |
| Cond-Mat 1993-2003 | 21363 | 91286 | 279 | 119.00 | 0.125 | 0.07 |
| Gnutella Aug 25 2002 | 22663 | 54693 | 66 | 28.58 | -0.173 | 0.04 |
| Internet | 22963 | 48436 | 2390 | 1085.20 | -0.198 | 0.08 |
| Thesaurus | 23132 | 297094 | 1062 | 1993.31 | -0.048 | 0.12 |
| Cora | 23166 | 89157 | 377 | 123.05 | -0.055 | 0.07 |
| AS Caida | 26475 | 53381 | 2628 | 1113.83 | -0.195 | 0.08 |
| Gnutella Aug 24 2002 | 26498 | 65359 | 355 | 35.03 | -0.008 | 0.04 |



\end{document}